\documentclass[reprint,
superscriptaddress,
amsmath,amssymb,
aps,prd,floatfix,
twocolumn,
showpacs,
]{revtex4-2}

\usepackage{amsthm}
\usepackage{tikz}
\usepackage{graphicx}
\usepackage{dcolumn}
\usepackage{bm}
\usepackage{braket}
\usepackage[english]{babel}
\usepackage{hyperref}
\usepackage{amsmath}
\usepackage{graphicx}
\usepackage{float}
\usepackage{siunitx}
\usepackage{xcolor}
\usepackage{xspace}
\usepackage{eucal}
\usepackage{ulem}
\usepackage{multirow}
\usepackage[utf8]{inputenc}
\usepackage{pgfplots}
\pgfplotsset{compat=1.14}
\usepackage{upgreek}
\usepackage{siunitx}
\usepackage{gensymb}
\usepackage{amsmath}
\usepackage[justification=justified,singlelinecheck=false]{caption}
\usepackage{subcaption}
\usepackage{textcomp}
\usepackage{comment}
\usepackage{cleveref}

\Crefname{figure}{Figure}{Figures}

\usepackage{todonotes}
\DeclareSIUnit\g{g}
\DeclareSIUnit\gal{Gal}
\DeclareSIUnit\torr{Torr}
\DeclareSIUnit\bar{Bar}
\DeclareSIUnit\kelvin{K}
\DeclareSIUnit\inch{inch}
\DeclareSIUnit\joule{J}
\DeclareSIUnit\rad{rad}

\begin{document}

\title{Quasi-monolithic heterodyne laser interferometer for inertial sensing}

\author{Yanqi Zhang}
\affiliation{Texas A\&M University, Aerospace Engineering \& Physics, 701 H.R. Bright Bldg., College Station, TX 77843}
\affiliation{Wyant College of Optical Sciences, The University of Arizona, 1630 E. University Blvd., Tucson, AZ 85721}
\author{Felipe Guzman}\email[Electronic mail: ]{felipe@tamu.edu}
\affiliation{Texas A\&M University, Aerospace Engineering \& Physics, 701 H.R. Bright Bldg., College Station, TX 77843}%

\date{\today} 

\begin{abstract}
We present a compact heterodyne laser interferometer developed for high-sensitivity displacement sensing applications. This interferometer consists of customized prisms and waveplates assembled as a quasi-monolithic unit to realize a miniaturized system. The interferometer design adopts a common-mode rejection scheme to provide a high rejection ratio to common environmental noise. Experimental tests in vacuum show a displacement sensitivity level of \SI{11}{pm/\sqrt{Hz}} at \SI{100}{mHz}, and as low as \SI{0.6}{pm/\sqrt{Hz}} above \SI{1}{Hz}. The prototype unit has the size of $\SI{20}{mm}\times\SI{20}{mm}\times\SI{10}{mm}$ and weighs \SI{4.5}{g}, which allows subsequent integration into compact systems.
\end{abstract}

\maketitle

\section{Introduction}
\label{sec:intro}
Displacement measuring interferometry (DMI) has long been an integral part of precision manufacturing and metrology engineering. Recent advancements in optomechanics, such as monolithic optomechanical inertial sensors\cite{Guzman2020,Hines:2020qdi}, extended the applications of DMI to the area of inertial sensing. Such optomechanical inertial sensors operating at sub-Hertz measurement bandwidths, consist of a mechanical resonator with a low resonant frequency of a few Hertz to increase the thermal-limited acceleration sensitivity, and an optical readout system to perform dynamic measurements of the test mass displacement.
The development of optical readout systems for low-frequency optomechanical inertial sensors faces several challenges. Firstly, the optical readout system is expected to achieve a high sensitivity in the millihertz frequency regime and simultaneously a large dynamic range to track the test mass motion. Due to the low resonant frequency and, therefore, the low stiffness of the mechanical resonator, the test mass motion may reach amplitudes of several microns to a few millimeters when subjected to large accelerations. Secondly, the footprint of the optical readout system should be comparable to the size of the mechanical resonator to facilitate system integration. Various types of DMI techniques exhibit a large dynamic range, but present other concerns at the same time. For example, the deep phase/frequency modulation interferometry (DPMI/DFMI) \cite{Heinzel:10} demonstrated displacement sensitivity of \SI{20}{pm/\sqrt{Hz}}, but requires complicated data processing algorithms for phase extraction. Furthermore, DFMI requires the use of lasers that are largely tunable in their frequency, which are typically complex, very expensive and result in additional complications to the system regarding their operation and frequency stability. In this paper, we propose an interferometer design based on heterodyne interferometry due to its inherent directional sensitivity and fast response speed.

\begin{figure}[htbp]
\centering
\fbox{\includegraphics[width=0.8\linewidth]{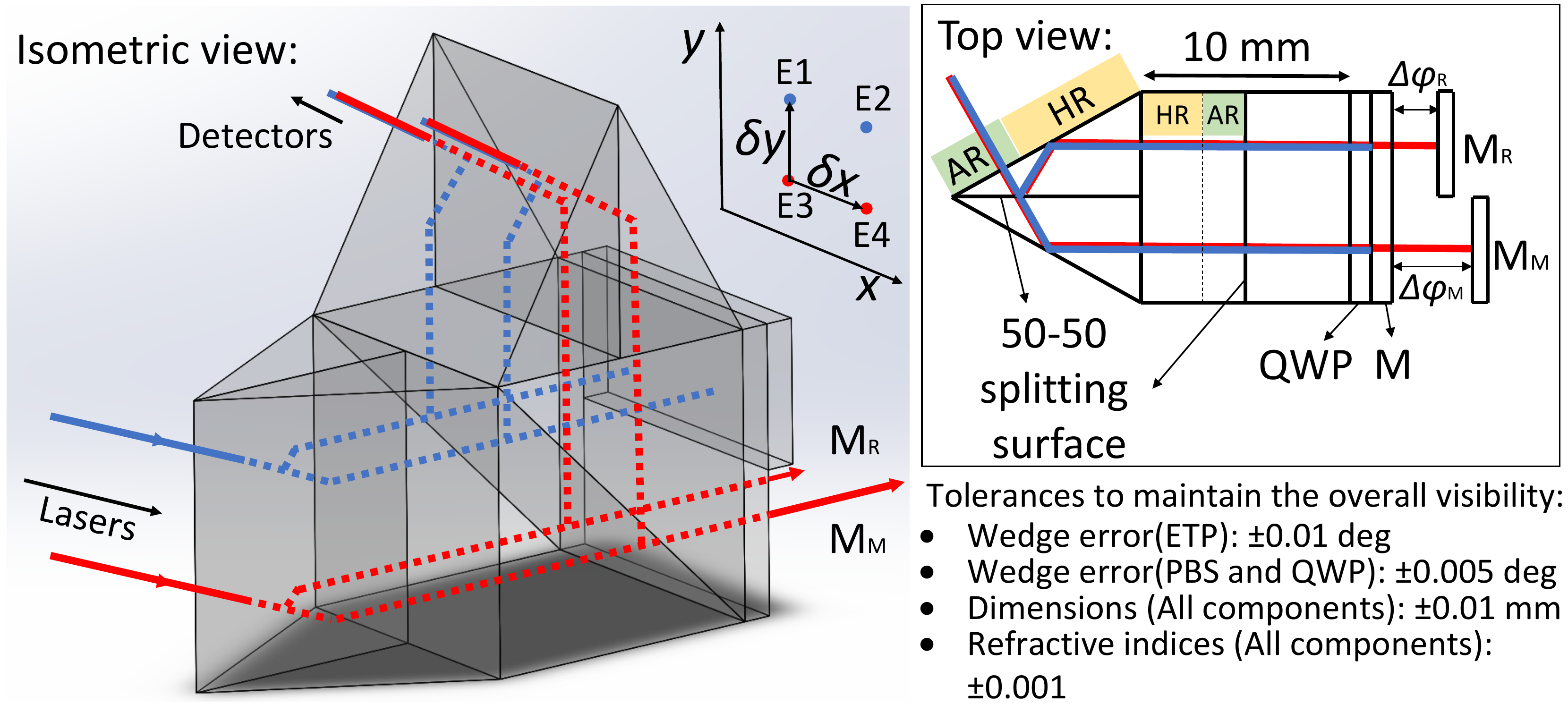}}
\caption{Schematic diagram of the quasi-monolithic interferometer unit in the isometric view and the top view. Two incoming laser beams are split into four beams by the 50-50 non-polarizing splitting surface embedded in the equilateral triangular prism (ETP). The beam pair with different frequencies constructs one interferometer. Manufacturing tolerances for each component to maintain the overall interferometer visibility over 90\% are listed in the figure.}
\label{fig:system-design}
\end{figure}

Efforts have been made to develop highly sensitive heterodyne displacement interferometers over the past decades using common-mode rejection schemes \cite{Heinzel:2004sr,Wand:2006AIPC,Guzman:2009thesis,Joo:2020JOSAA,Zhang:2021}. In the Laser Interferometer Space Antenna (LISA) \cite{Heinzel:2004sr} and its demonstration unit, LISA Pathfinder (LPF) \cite{Wand:2006AIPC, Guzman:2009thesis}, four heterodyne interferometers are integrated onto one bench as the optical readout system. It achieves a sensitivity level of $\SI{10}{pm/\sqrt{Hz}}$ at $\SI{1}{mHz}$ on ground, and $\SI{30}{fm/\sqrt{Hz}}$ in space when measuring the displacement of a free-falling test mass. However, the manufacturing of this interferometer unit requires complicated alignment and bonding techniques, which makes it difficult to replicate and deploy in other sensing applications. Beyond space-based gravitational-wave observation, several other applications can benefit from such compact high-sensitivity laser interferometers such as spacecraft drag-free control\,\cite{GUILHERME2008365,DEROSA2011394} or the characterization and calibration of laser direct writing lithography\,\cite{Jaramillo2017}, which require precisions at or below the nanometer level, as well as dimensional metrology, and machining tools in manufacturing and precision measurements, striving precisions from picometers up to microns\,\cite{HANSEN2006721}.

Other types of common-mode heterodyne interferometers \cite{Joo:2020JOSAA,Zhang:2021} utilize a reference interferometer to monitor environmental noise and reached a picometer-level sensitivity at \SI{100}{mHz}. However, the prototype systems have large footprints and require alignment of individual components, making them less appealing in portable applications. 

To this end and building upon the expertise gained from the LPF interferometer development, we propose a novel configuration of a highly sensitive and compact optical assembly that comprises two heterodyne interferometers operating at normal incidence. Upon further developments and benefiting from the design and operational principles of LPF, this results in a miniaturized quasi-monolithic unit with a high common-mode noise rejection. The proposed interferometer consists of a quasi-monolithic assembly of the optical components such as prisms and waveplates. The benchtop prototype achieves a displacement sensitivity level of \SI{0.6}{pm/\sqrt{Hz}} at \SI{1}{Hz} when tested in vacuum. The assembly unit fits in a \SI{20}{mm} cube and weighs \SI{4.5}{g}, which makes it compatible with monolithically-fabricated mechanical resonators \cite{Hines:2020qdi}, and highly compact instruments. 

\section{System design}
\label{sec:design}

The quasi-monolithic interferometer unit consists of two specially designed equilateral triangular prisms (ETP), a polarizing beam splitter (PBS), a quarter waveplate (QWP), and a mirror $M$, as depicted in Figure~\ref{fig:system-design}. A 50-50 non-polarizing beam splitting surface is embedded in the middle of the ETP, dividing the ETP into two 30-60-90 prisms. The optical coatings on the side surface of the ETP are split into two parts, where the part closed to the 30-degree angle is treated with an anti-reflection (AR) coating and the other part has a high-reflectivity (HR) coating. Total internal reflection (TIR) is avoided in this assembly to maintain the polarization state of the incoming laser beams. The mirror $M$ can be replaced with the HR coating to further reduce the system footprint. 

After passing through two individual acousto-optical modulators (AOM), two laser beams are shifted to slightly different frequencies and enter the side surface of the ETP at normal incidence. These two beams are split into four after transversing through the 50-50 splitting surface. All four beams are reflected by the HR coatings on the ETP and enter the PBS perpendicularly. On the measurement end, the top two beams are reflected by the attached mirror $M$ that serves as the global reference. One of the bottom beams is reflected by the local reference mirror $M_\mathrm{R}$ that is fixed near the target, while the other bottom beam is incident on the mirror $M_\mathrm{M}$ that is mounted on the target. The reflected beams return to the PBS again and propagate towards the second ETP, where the beam pairs $E_1 \& E_3$ and $E_2 \& E_4$ overlap to build the reference interferometer (RIFO) and the measurement interferometer (MIFO), respectively, and are detected by their corresponding photodetectors (PD). 

The electric-field amplitude of the four beams at the detection plane can be expressed as 
\begin{align}
E_1 &= E_{10}\exp{[i(\omega+\Delta \omega_1)t+\phi_0(0,\delta y)]},\\
E_2 &= E_{20}\exp{[i(\omega+\Delta \omega_1)t+\phi_0(\delta x,\delta y)]},\\
E_3 &= E_{30} \exp{[i(\omega+\Delta \omega_2)t+\phi_0(0,0)+\Delta \phi_\mathrm{R}]},\\
E_4 &= E_{40} \exp{[i(\omega+\Delta \omega_2)t+\phi_0(\delta x,0)+\Delta \phi_\mathrm{M}]},
\end{align}
where $\phi_0(x,y)$ represents the phase shift of each beam when propagating from the entry plane to the detection plane. The terms $\Delta \phi_\mathrm{R}$ and $\Delta \phi_\mathrm{M}$ denote the phase contribution from the optical pathlength difference (OPD) between mirrors $M$ and $M_\mathrm{R}$, and between mirrors $M$ and $M_\mathrm{M}$. The detected irradiance signals of MIFO and RIFO are expressed by 
\begin{align}
    I_\mathrm{M} &= I_\mathrm{M0}[1-V_1 \cos(\Delta \omega t +\phi_0(0,\delta y)+\Delta \phi_\mathrm{M})],\\
    I_\mathrm{R} &= I_\mathrm{R0}[1-V_2 \cos({\Delta \omega t +\phi_0(0,\delta y) + \Delta \phi_\mathrm{R})}],
\end{align}
where $V$ is the interferometer visibility, $\Delta \omega=\Delta \omega_2-\Delta \omega_1$ is the heterodyne frequency, and $\phi_0(0,\delta y)=\phi_0(x,0)-\phi_0(x,\delta y)$, is the phase difference between two vertically aligned beams due to any potential structural instability along the y-direction. By performing the differential operation between the phase readouts from MIFO and RIFO, the y-axis structural fluctuation terms $\phi_0(0,\delta y)$ cancel out, enhancing the overall system sensitivity. The target displacement $d$ is calculated as 
\begin{equation}\label{eq:disp}
    d=\frac{\Delta \phi }{2 \cdot 2\pi} \lambda = \frac{\Delta \phi_M-\Delta \phi_R}{2 \cdot 2\pi} \lambda,
\end{equation}
where $\lambda$ is the laser wavelength.  

The spatially separated beams in the design mitigate the inherent periodic errors in the heterodyne interferometer \cite{Wu:1998ao,Joo:10,Ellis:11}. Due to the compact size of the assembly, the optical paths between MIFO and RIFO are highly common, providing a high rejection ratio to common noise sources such as temperature fluctuations. The manufacturing errors on the prisms, such as dimensions and wedge angles, can be compensated by adjusting the positions and incident angles of the input beams. We conducted a Monte Carlo analysis on the dimensions and geometry of the components to maintain the overall interferometer visibilities, $V$, above 90\% at a 90\% yield, and determine the tolerances (see Figure~\ref{fig:system-design}) for the optical components. In this analysis, we considered fiber injectors with a limited resolution of \SI{0.01}{mm} for the displacement and \SI{0.01}{degree} as compensators for the angle of incidence.

\section{Prototype characterization}

\begin{figure}[t]
\centering
\fbox{\includegraphics[width=0.8\linewidth]{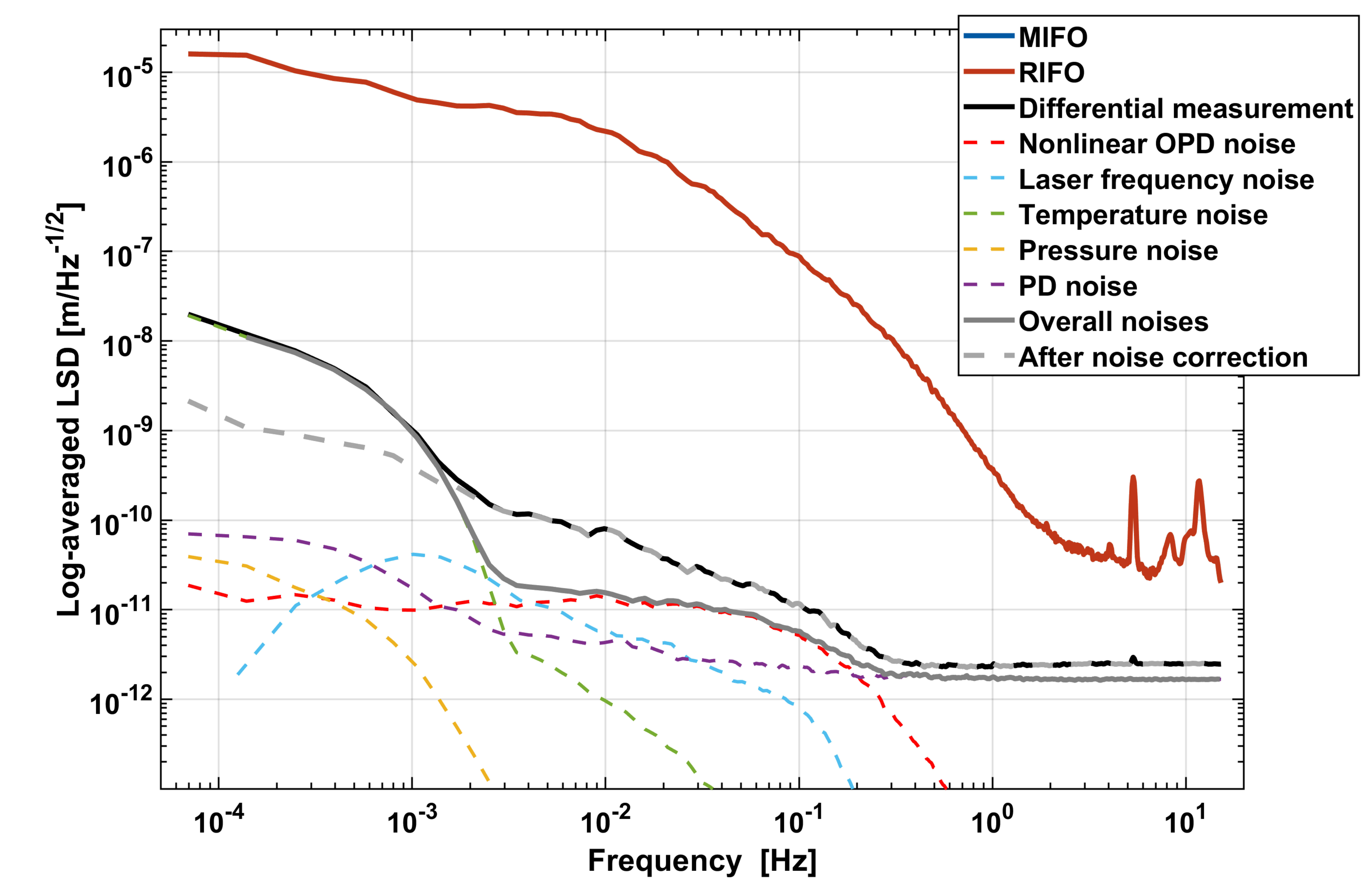}}
\caption{Logarithmic-averaged linear spectral densities (LSD) of 4-hour measurements from individual interferometers MIFO and RIFO, and the differential measurement, as well as the contributions from different noise sources. The sensitivity levels of the individual interferometers are both \SI{9.25e-8}{m/\sqrt{Hz}} at \SI{100}{mHz}, and overall systematic sensitivity is enhanced to \SI{1.16e-11}{m/\sqrt{Hz}} at \SI{100}{mHz} by the differential operation.}
\label{fig:lsd}
\end{figure}

We developed a prototype interferometer according to the design shown in Figure~\ref{fig:system-design}, made of N-BK7 with coatings for a wavelength of \SI{1550}{nm}. Individual components are cemented by UV curing using the adhesive NOA61 for index matching. The relative positions and orientation of two beam injectors can be easily adjusted during the alignment process with respect to the interferometer assembly, to maximize the visibility of the two interferometers.
The prototype is tested inside a vacuum chamber that operates at \SI{4}{\milli \Pa}. In this preliminary test, one static mirror serves as both the measurement and the reference mirror to characterize the interferometric system noise only. Figure~\ref{fig:lsd} shows the logarithmic-averaged linear spectral density (LSD) of a 4-hour measurement for the individual interferometers MIFO and RIFO, and the differential displacement calculated by Equation~\ref{eq:disp}. The LSD of MIFO and RIFO measurements highly overlap due to the common paths between the two interferometers. The logarithmic-averaged traces shows that MIFO and RIFO achieve the sensitivity level of \SI{9.25e-8}{m/\sqrt{Hz}} at \SI{100}{mHz} frequency, while the differential measurement enhances the sensitivity by over two orders of magnitude, reaching \SI{1.16e-11}{m/\sqrt{Hz}} at \SI{100}{mHz}. 

We investigated various noise sources, including temperature fluctuations, nonlinear OPD noise, laser frequency noise, pressure fluctuations, and detection system noise, to estimate their contribution to the displacement measurement, as shown in the dashed lines in Figure~\ref{fig:lsd}. 

\begin{figure}[t]
\centering
\fbox{\includegraphics[width=0.8\linewidth]{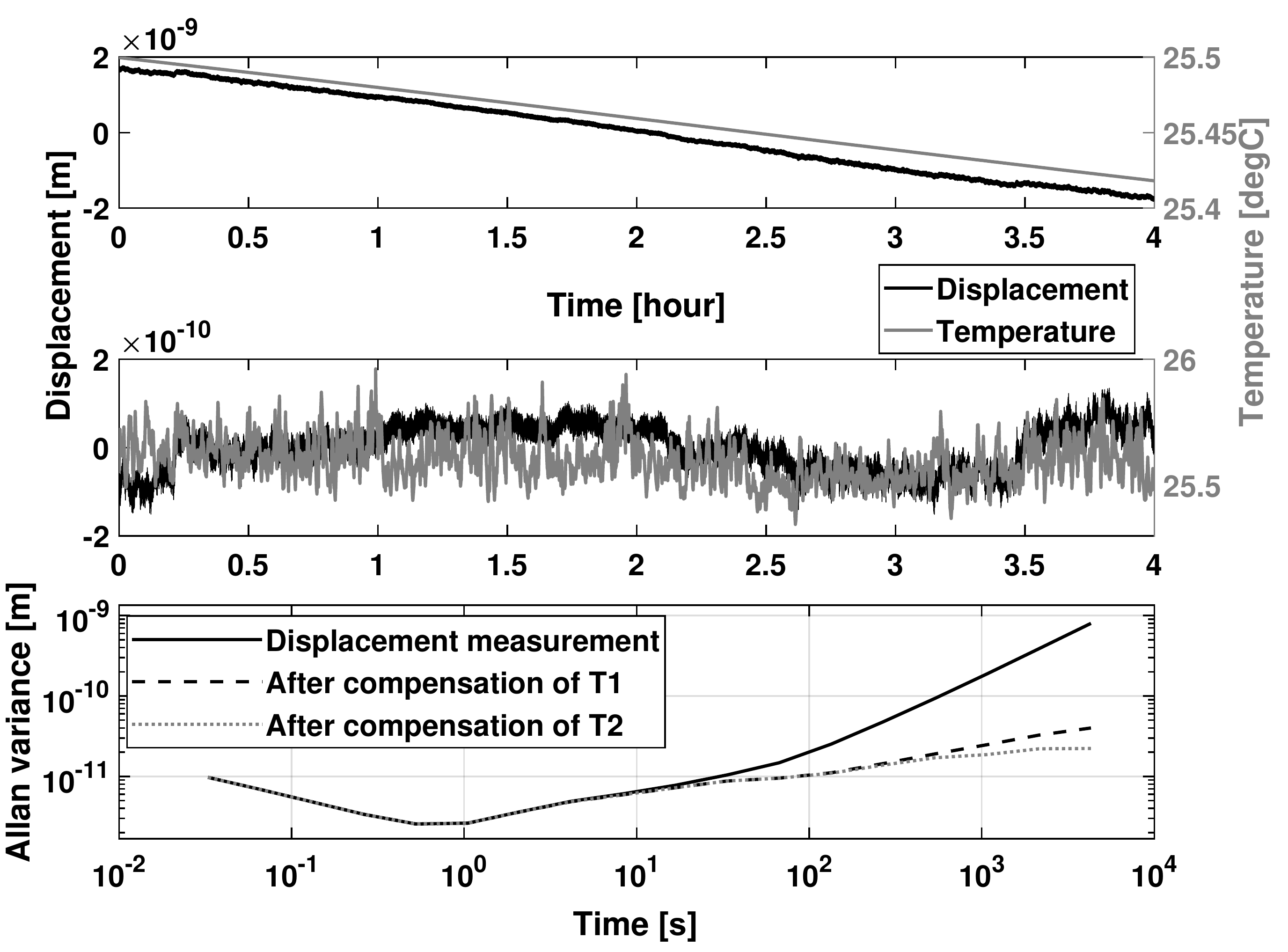}}
\caption{(a) The original displacement measurement and the temperature measurement $T1$ with the temperature sensor installed inside the chamber; (b) The displacement after compensating the thermal drift using $T1$, and the temperature measurement $T2$ with the temperature sensor installed outside the chamber; (c) Allan variances of the original displacement measurement, and the displacement after compensating for $T1$ and $T2$ measurements, respectively.}
\label{fig:corr}
\end{figure}

Two thermistors are used as temperature sensors. One of them is installed inside the vacuum chamber, attached to the breadboard, and the other one is outside the chamber in air, to measure temperature fluctuations at a sampling frequency of \SI{1}{Hz}. Figure~\ref{fig:corr}(a) shows the time series of the 4-hour measurements of the differential displacement and the temperature measurement ($T1$) inside the chamber. The displacement measurement is corrected for the thermal drift using the $T1$ measurement, resulting in the trace shown in Figure~\ref{fig:corr}(b) along with the temperature measurement ($T2$) outside the chamber. Figure~\ref{fig:corr}(c) shows the Allan deviation of the differential displacement to be \SI{1.8e-11}{m} over an integration time of \SI{1000}{s} after correction. The temperature coupling coefficients for $T1$ and $T2$ are estimated to be \SI{4.41e-8}{m/K} and \SI{6.64e-10}{m/K}.

The electromagnetic interference of the radio-frequency (RF) signals that drive the two AOMs imprints spurious sidebands onto the laser beams, resulting in a ghost signal with the same heterodyne frequency $f_\mathrm{het}$. The phase error introduced by this parasitic signal is referred to as the nonlinear OPD noise, the amplitude of which is related to the phase readouts of the individual interferometers $\Delta \phi_\mathrm{M}$ and $\Delta \phi_\mathrm{R}$. By applying a linear fitting algorithm described in \cite{Zhang:2021}, we can separate this noise contribution from the displacement measurement. In the experimental setup, this noise effect is mitigated by RF shielding. 

The OPD between the two interferometer arms determines the coupling coefficient of the laser frequency noise. The proposed quasi-monolithic interferometer is designed to have a symmetric configuration that helps mitigate the effects of laser frequency noise. Moreover, in the test setup where only one static mirror is applied for both measurement and reference mirrors, the OPD is further reduced between the two individual interferometers. In the test setup, the laser frequency noise is monitored using a self-referenced fiber delay-line interferometer (DIFO) \cite{Zhang:2021} that amplifies the laser frequency noise effects using the intentional unequal arm lengths. The laser frequency noise is extracted from the phase measurement of DIFO, which is bandpass filtered between \SI{1}{mHz} and \SI{100}{mHz} to mitigate the contributions of other noise sources \cite{Zhang:2021}. 

We recorded the displacement measurements when the vacuum pumps were turned off to prevent mechanical vibrations from coupling into the phase measurement. In this case, the pressure rises inevitably inside the chamber and is monitored by the pressure gauge at a sampling frequency of \SI{1}{Hz}. The coupling coefficient of the pressure variation is estimated to be \SI{2.1e-12}{m/Pa} based on a measurement of the pump-down process. 

The detection system noise such as the analog-to-digital conversion (ADC), photodetectors~\cite{GuzmanCervantes:2011zz}, and phasemeter noise~\cite{Gerberding2015} determine the lowest noise floor that an interferometric system can achieve. A zero-test is performed for both PDs, where the detected irradiance signal from a single PD is split into two input channels of the phasemeter. 

\begin{figure}[htbp]
\centering
\fbox{\includegraphics[width=0.8\linewidth]{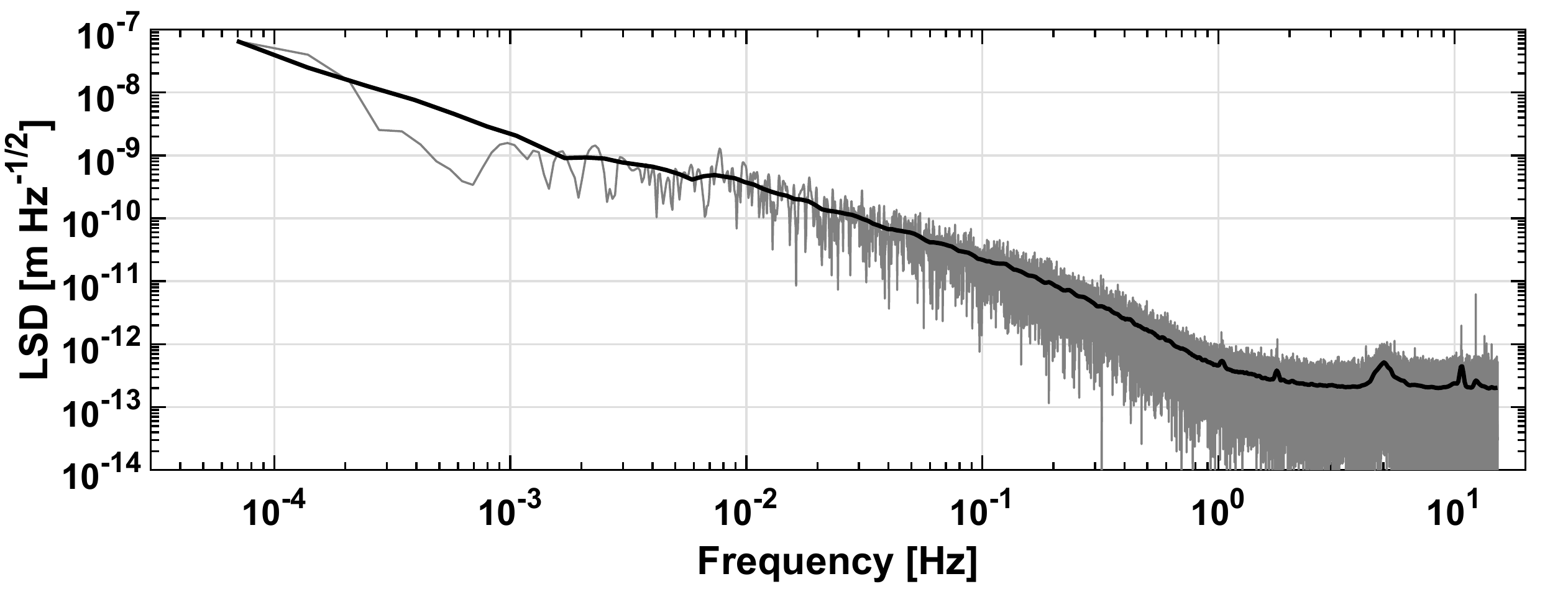}}
\caption{The LSD of a 4-hour differential measurement with the improved PD system performance. The sensitivity floor of this differential measurement is \SI{0.6}{pm/\sqrt{Hz}} above \SI{1}{Hz}.}
\label{fig:lsd-adev-m2}
\end{figure}

In the low-frequency regime below \SI{1}{mHz}, the temperature fluctuations are the dominant noise sources. In the frequency bandwidth above \SI{1}{Hz}, the interferometer performance is limited by the PD noise,which can be further improved by increasing the optical power incident on the photodetectors, as well as by improving the interferometer alignment for a better fringe contrast. We took another 4-hour measurement with these improvements under the same environmental conditions, as shown in Figure~\ref{fig:lsd-adev-m2}. This plot shows that for the differential measurement, the LSD is \SI{6e-13}{m/\sqrt{Hz}} at \SI{1}{Hz}.

Laser frequency noise and temperature fluctuations are likely the main factors limiting the interferometer sensitivity at low frequencies due to the fact that we are operating a free-running laser, and our fiber-based DIFO is not good sensor of laser frequency fluctuations over this bandwidth. Furthermore, when using a free-running laser, this will likely become a significant noise source when the axial distance between the reference mirror $M_\mathrm{R}$ and the measurement mirror $M_\mathrm{M}$ increases. This can be mitigated by adopting laser frequency stabilization systems with active feedback control. Since the test mirrors, the optical assembly, and the fiber injectors are all currently supported with off-the-shelf kinematic mounts, we consider it plausible that the residual noise floor and its rise toward lower frequencies originate in their inherent mechanical and material instabilities~\cite{ellisphd}.

\begin{figure}[htbp]
\centering
\fbox{\includegraphics[width=0.8\linewidth]{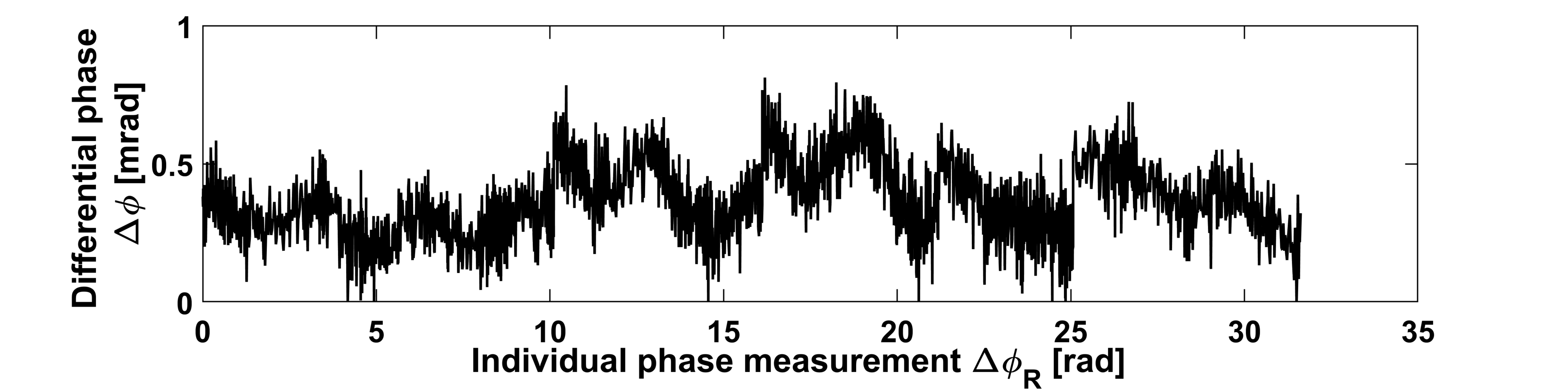}}
\caption{The differential phase readout $\Delta\phi$ with respect to one of the individual phase readout $\Delta\phi_\mathrm{R}$.}
\label{fig:periodicity}
\end{figure}

We have design this interferometer assembly to mitigate inherent periodic errors. Any stray light originating at the optical interfaces due to normal incidence would typically result in a periodic error pattern in the interferometer signal that can be easily recognized by plotting the differential phase between the two interferometers with respect to a single interferometer phase~\cite{Wand:2006CQG}. Figure \ref{fig:periodicity} shows the differential phase $\Delta\phi$ versus one of the individual phase readout $\Delta\phi_\mathrm{R}$ over a few interferometer fringes (10\,rad). It can be clearly observed that no consistent periodic pattern is recognizable in this plot, where we conclude that any effect from potential stray light is negligible.

\begin{figure}[t]
\centering
\fbox{\includegraphics[width=0.6\linewidth]{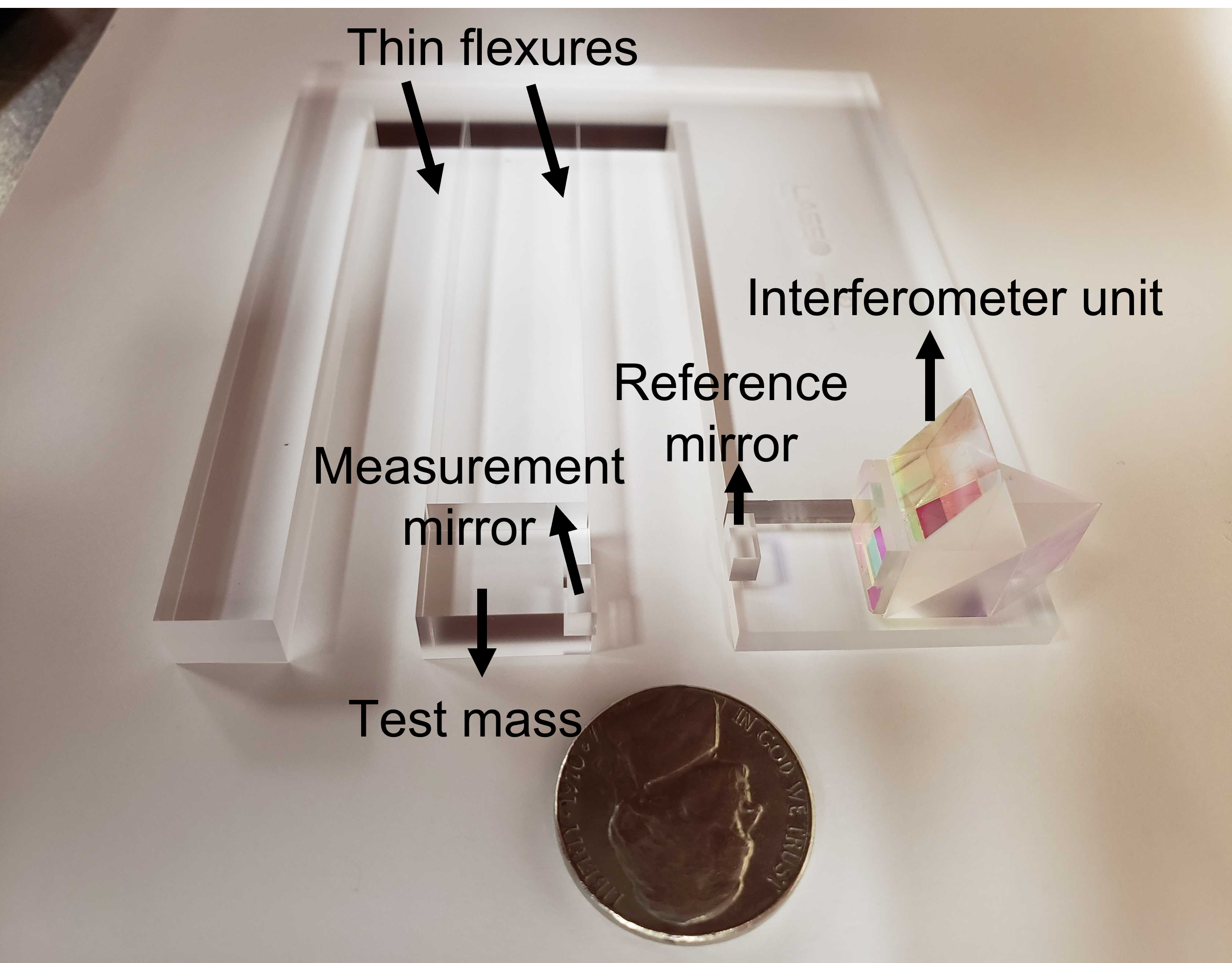}}
\caption{System assembly by integrating the proposed quasi-monolithic interferometer as the optical readout system with the monolithically fabricated fused-silica mechanical resonator \cite{Hines:2020qdi}, to construct a portable and highly sensitive optomechanical inertial sensor. The overall footprint is compared with a nickel coin in the picture.}
\label{fig:assembly}
\end{figure}

Figure \ref{fig:assembly} shows the system assembly where the interferometer unit is integrated as the optical readout system for an optomechanical inertial sensor. A notch is etched on the test mass of the mechanical resonator to mount the mirror $M_\mathrm{M}$. The mirror $M_\mathrm{R}$ is mounted on the baseplate of the resonator. The overall footprint of the inertial sensor is $\SI{90}{mm} \times \SI{80}{mm} \times \SI{20}{mm}$, providing a compact solution to highly sensitive accelerometers. 

\section{Summary and outlook}

In this paper, we presented the design of a quasi-monolithic heterodyne laser interferometer assembly that utilizes a common-mode rejection scheme to enhance the overall sensitivity while maintaining a very compact form factor. Preliminary tests in our lab shows a sensitivity level of \SI{0.6}{pm/\sqrt{Hz}} above \SI{1}{Hz} in vacuum, which is greatly enhanced from the two individual heterodyne interferometers.

Upon integration of the interferometer unit with the mechanical resonator, upcoming results will involve true acceleration measurements at low frequencies using the fully assembled optomechanical inertial sensing units. Such a unit will also include improvements such as a laser frequency stabilization using a fiber-coupled HCN gas cell and lower-noise photoreceivers. Such a laser frequency stabilization system is under development in our lab utilizing the HCN gas cell to establish the molecular spectroscopy locking of the laser frequency.

\subsection*{Funding} 
National Geospatial-Intelligence Agency (NGA) through grant HMA04762010016; National Science Foundation (NSF) through grants PHY-2045579 and ECCS-1945832; National Aeronautics and Space Administration (NASA) through grant 80NSSC20K1723.

\subsection*{Acknowledgments}
The authors thank Adam Hines, Bo Stoddart, and Lee Ann Capistran for implementing the data acquisition script for temperature and pressure measurements. The authors also thank Dr. Jose Sanjuan for proofreading the manuscript. 

\subsection*{Disclosures}
The authors declare no conflicts of interest.

\subsection*{Data availability}
Data underlying the results presented in this paper are not publicly available at this time but may be obtained from the authors upon reasonable request.

\bibliography{References}

\end{document}